# Estimation and Sensitivity Analysis for Causal Decomposition in Health Disparity Research

Soojin Park[1], Xu Qin[2], and Chioun Lee[1]

[1]University of California, Riverside

[2]University of Pittsburgh

Abstract

In the field of disparities research, there has been growing interest in developing a counterfactual-based decomposition analysis to identify underlying mediating mechanisms that help reduce disparities in populations. Despite rapid development in the area, most prior studies have been limited to regression-based methods, undermining the possibility of addressing complex models with multiple mediators and/or heterogeneous effects. We propose a novel estimation method that effectively addresses complex models. Moreover, we develop a sensitivity analysis for possible violations of an identification assumption. The proposed method and sensitivity analysis are demonstrated with data from the Midlife Development in the US study to investigate the degree to which disparities in cardiovascular health at the intersection of race and gender would be reduced if the distributions of education and perceived discrimination were the same across intersectional groups.

Estimation and Sensitivity Analysis for Causal Decomposition in Health Disparity Research

## 1. Introduction

Despite gradual declines in cardiovascular disease (CVD) mortality in the US over 50 years, racial differences in the burden of CVD continue to play a substantial role in maintaining racial differences in life expectancy (Carnethon et al., 2017; Leigh, Alvarez, & Rodriguez, 2016). In 2010, the American Heart Association (AHA) introduced a new metric, "ideal cardiovascular health," to improve cardiovascular health and reduce health disparities in populations (Lloyd-Jones et al., 2010). Yet, cumulative evidence shows that African Americans (hereafter "Blacks") have worse cardiovascular health than non-Hispanic Whites (hereafter "Whites") and that such a racial gap appears larger for women than men (Pool, Ning, Lloyd-Jones, & Allen, 2017), indicating that populations that fall into multiple minority statuses (Black women) are particularly vulnerable to poor cardiovascular health. In order to identify mediating factors that potentially reduce disparities, we investigated hypothetical interventions that would simultaneously equalize the distributions between non-marginalized groups (e.g., White men) and marginalized groups (e.g., Black women) of two well-established psychosocial mediators: education and perceived discrimination.

The current study is motivated by three methodological challenges to investigating multiple mediating mechanisms underlying health disparities across different race-gender groups. First, investigators who have used multiple mediators are often concerned about the causal ordering of mediators, particularly when their temporal ordering is unclear. For example, one could reasonably speculate that a low level of education could exacerbate discrimination in addition to its adverse effect on cardiovascular health. On the contrary, one could imagine that perceived discrimination could lead to less education. A method that does not address this bidirectional relationship between mediators often fails to capture an accurate effect of the mediators in reducing health disparities. Second, the causal structural model underlying the identification of mediators is complex in many circumstances; for example, Bauer and Scheim (2019) have shown differential effects. That is, the effect of perceived discrimination on psychological distress varies by race-gender group. Thus, it is essential to use an estimator that addresses heterogeneous effects. Third, because results may not be valid if one of the identification assumptions for the effects of interest is violated, sensitivity analysis for possible violations of assumptions is needed. The goals of the current study, therefore, are to 1) review a method that can address multiple mediators in which the causal ordering of the mediators is unclear, 2) develop a novel estimator based on inverse-propensity scores and imputation that can address heterogeneous effects and/or multiple mediators, and 3) develop a sensitivity analysis against possible violations of an identification assumption.

This paper proceeds as follows. Section 2 reviews causal decomposition analysis in the context of health disparities research, and Section 3 presents the identification assumptions and results. Section 4 presents our estimation method in comparison to regression-based methods, and Section 5 presents our sensitivity analysis. In Section 6, we demonstrate our estimation method and sensitivity analysis using data from the Midlife Development in the US (MIDUS) study. Finally, Section 7 discusses the implications of the study for best practices for disparity research. The R code used for our case study is given in appendix D. Open-source software for R (*causal.decomp*) that implements the proposed estimation method is available from https://cran.r-project.org/web/packages/causal.decomp/index.html.

## 2. Causal Decomposition Analysis: A Review

### 2.1. Observed Disparities across Intersectional Groups

Intersectionality is a theoretical framework that views multiple categories (such as gender and race) as interacting in a matrix of domination, producing distinct inequalities with adverse outcomes for marginalized groups (Collins, 1990). For example, suppose that we investigate disparities in cardiovascular health at the intersection of gender and race. Following the intercategorical approach used in Bauer and Scheim (2019), the nexus of self-identified race and gender implies four intersectional groups: White men, White women, Black men, and Black women. Intersectionality theory suggests that women of color (Black women) will have poorer cardiovascular health than other race-gender groups.

The issue is that the causal effect on cardiovascular health of race-gender intersectional status is hard to estimate because 1) race or gender is difficult to define precisely, and 2) race is not randomized. VanderWeele and Robinson (2014) argued that the effect of race may indicate, separately or jointly, the effect of genetic background, physical features (e.g., skin color and its perception of others), parents' physical features, or cultural context. Moreover, health is determined by both sex and gender. "Sex" refers to biological aspects of maleness and femaleness, while "gender" refers to socially constructed and enacted roles and behaviors which arise in a historical and cultural context (NIH, 2020). In most population-based studies, respondents were not asked about their sex (female vs. male) and gender identity (women vs. men) separately. In our study, we assume that self-identified sex reflects gender identity and use race to encompass all of the aforementioned components (i.e., genetic background, physical features, parental physical features, and cultural context).

Even when race and gender are precisely defined, race is not randomized, so its effect would be correlated with confounding variables, such as genetic vulnerability, family socioeconomic status (SES), neighborhood SES, etc. For identifying a causal effect of race, a randomized intervention will be required. However, socially defined characteristics such as race are not something we can intervene on. For example, we cannot intervene to make a Black individual White and compare their counterfactual outcomes, except for some instances (e.g., randomizing the employer's perception of an applicant's race). Therefore, VanderWeele and Robinson (2014) suggested estimating the observed disparity between the non-marginalized and marginalized groups by simply comparing means. Because of lack of randomization, such a disparity comparison no longer corresponds with a causal effect of race and is considered associational (VanderWeele & Hernán, 2012). Jackson (2017) expanded this idea to consider intersectional disparities as associative quantities. We adopt this approach and focus on the observed disparity in cardiovascular health between non-marginalized and marginalized groups.

## 2.2. Effect of Mediating Variables

Simply observing health disparities between non-marginalized and marginalized groups does not necessarily explain why the disparities exist or how to reduce them (Jackson, 2017), but investigating mediating mechanisms can help to inform policy interventions that reduce health disparities. Although we avoid causal interpretations of socially defined characteristics, this does not prevent us from making causal interpretations of their mediation effects. Estimating natural direct and indirect effects (Pearl, 2001; Robins & Greenland, 1992) defined under the potential-outcome framework has been an integral part of the history of causal mediation analysis. Jackson (2017) extended the approach of interventional effects to the setting of intersectional disparity. Bauer and Scheim (2019) also adopted this approach and applied VanderWeele's 3-way decomposition analysis in the context of health disparities at the intersection of gender, sexuality, and race/ethnicity. However, estimating natural effects can be challenging in disparities research. Natural indirect effects for Black women are defined, for example, as the expected change in the outcome observed after fixing the intersectional status to Black women but varying the level of discrimination to the value that would have resulted from being born a Black woman versus a White man. Estimating natural effects requires setting the mediator value (discrimination) for each Black woman to a potential value that would have resulted had she been born a White man. Considering this counterfactual outcome is in some sense meaningless since race and gender are essentially non-modifiable.

In response to this concern, VanderWeele and Robinson (2014) considered the use of *randomized interventional analogues* of natural direct and indirect effects (hereafter, interventional effects) in the context of health disparities. This intervention requires setting the level of discrimination of Black women to the level of White men as a group. This requirement is more sensible than setting the mediator value (discrimination) for each Black woman to a potential value that would have resulted had she been born a White man, as natural effects require. Moreover, interventional effects allow us to ask policy-relevant questions, such as *how much disparity would be reduced if we decrease the level of discrimination of Black women to the level of White men?*

Methodologically, interventional indirect effects require weaker assumptions than natural indirect effects. Identifying natural indirect effects requires three assumptions: 1) exposure-outcome or exposure-mediator confounders are measured, 2) mediator-outcome confounders are measured, and 3) the exposure does not affect the mediator-outcome confounders. The first assumption is not required in the context of our study because no causal effect of intersectional status is specified, and we only focus on the observed disparity. The second and third assumptions are strong, but the latter is particularly strong given that health disparities can be determined by a myriad of factors throughout one's life. Interventional effects do not require this third assumption. The second assumption alone allows the identification of path-specific effects involving multiple mediators as long as the causal ordering between the mediators is determined, as shown in Jackson (2018).

Despite methodological developments in recent years, prior studies have largely relied on a regression-based estimator (e.g., Jackson & VanderWeele, 2018; VanderWeele & Robinson, 2014). Regression is a suitable method if the causal structural model is simple with a single mediator and no differential effects. Recently, Jackson (2020) proposed a weighting method, using ratio-of-mediator probability and inverse odds ratios, that handles models with differential effects and a single mediator effectively. Therefore, as a next step, it is essential to develop an estimator that can address complex models with differential effects and multiple mediators in the context of disparity research.

In addition to employing a flexible estimator, an important problem is that results may be sensitive to a possible violation of assumptions invoked for the identification of the effects of interest. Many sensitivity analysis techniques have been developed to evaluate possible violations of confounding assumptions in the context of causal mediation studies based on natural indirect effects (e.g., Hong, Qin, & Yang, 2018; VanderWeele, 2010a; Imai, Keele, & Yamamoto, 2010; VanderWeele & Chiba, 2014; Imai & Yamamoto, 2013). Sensitivity analysis is an essential part of causal mediation studies but it has not yet appeared in the disparity literature. This is perhaps because the existing sensitivity analysis techniques based on natural indirect effects have not been extended to randomized interventional effects, which are often employed in disparity research.

### 3. Identification

In this section, we formally define the effects of our interest and provide identification results. As discussed in the introduction, the causal ordering between the mediators (i.e., perceived discrimination and education) is arguable and, without knowing the causal ordering, it is impossible to identify and estimate the interventional effects via each mediator. Therefore, we review a method of considering mediators jointly, which is previously proposed by VanderWeele and Vansteelandt (2014) in the context of natural indirect effects.

#### 3.1. Notation and Definitions

Figure 1 depicts a directed acyclic graph (DAG) that represents causal structural models. We denote the intersectional status $R$ and cardiovascular health $Y$. Following Kaufman (2008), we assume complex historical structures responsible for racism and sexism ($H$) give rise to an association between the intersectional status ($R$), childhood SES ($X_1$), and genetic vulnerability ($C_2$). We also assume that age ($C_1$) is correlated with history. There are three intermediate variables between the intersectional status and cardiovascular health: child abuse ($X_2$), perceived discrimination ($D$), and education ($M$). Here, perceived discrimination and education serve as mediators of our interest. For notational simplicity, the vector of baseline covariates consisting of $C_1$ and $C_2$ is denoted as

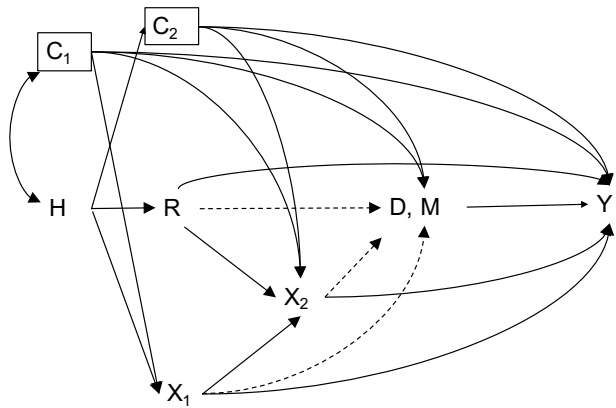

*Figure 1*. Directed acyclic graph showing the relationship between intersectional status, cardiovascular health, and three potential mediators

Note. 1) Diagram represents the relationship between race and gender intersectional status $R$, cardiovascular health $Y$, perceived discrimination $D$, and education $M$, as well as history $H$, age $C_1$, genetic vulnerability $C_2$, childhood SES $X_1$, and child abuse $X_2$. 2) Solid lines represent relationships that are preserved and dashed lines represent relationships that are removed by intervening on $D$ and $M$. 3) Placing a box around the conditioning variables implies that a disparity is considered within levels of these variables.

**C**. The vector of intermediate confounders measured at the same time as the exposure ($X_1$) and measured after the exposure ($X_2$) is denoted as **X**. The supports of these distributions are $\mathcal{C}$ and $\mathcal{X}$, respectively.

We begin by defining the initial disparity. Let the intersectional group indicator consists of four groups: White men ($R = 0$), White women ($R = 1$), Black men ($R = 2$), and Black women ($R = 3$). Suppose that White men ($R = 0$) is the reference group and that the rest of the groups are the comparison groups. Then, the initial disparity ($\tau$) is defined as the difference in the cardiovascular health outcome between a comparison group ($R = r$) and the reference group ($R = 0$) given baseline covariates, which are age ($C_1$) and genetic vulnerability ($C_2$). These covariates affect cardiovascular health and are associated with intersectional status through complex historical processes responsible for gender and race differences. We consider disparities within the same age and genetic vulnerability levels because these covariates are given and not manipulable. Formally, the initial disparity is defined as $\tau(r, 0) \equiv \sum_{\mathbf{c}} E[Y|R = r, \mathbf{c}]P(\mathbf{c}) - E[Y|R = 0, \mathbf{c}]P(\mathbf{c})$, where $r \in \{1, 2, 3\}$ and $\mathbf{c} \in \mathcal{C}$. Since no causal interpretation is given to this estimand ($\tau$), no causal identification assumptions are required. However, positivity (i.e., $P(R = r|\mathbf{c}) > 0$) is required for a nonparametric identification.

Our main interest is how much the initial disparity would be reduced or remain if we intervened so that the distributions of education and perceived discrimination were the same between the reference group and a comparison group. Under the stable-unit-treatment-value assumption (SUTVA[1]), let $G_{d|\mathbf{c}}(0)$ and $G_{m|\mathbf{c}}(0)$ be random draws, given baseline covariates ($\mathbf{C} = \mathbf{c}$), from the distributions of the mediators $D$ and $M$, respectively, under the reference group ($R = 0$). As a result, $E[Y(G_{d|\mathbf{c}}(0), G_{m|\mathbf{c}}(0))|R = r]$ is the expected cardiovascular health for a comparison group ($R = r$) that would have been observed if perceived discrimination and education were randomly drawn from the distributions of these mediators for members of the reference group who have the same age and same genetic vulnerability. Using this counterfactual quantity, the disparity reduction ($\delta$) and disparity remaining ($\zeta$) are defined as

$$
\begin{aligned}
\delta(r) &\equiv \sum_{\mathbf{c}} E[Y|R = r, \mathbf{c}]P(\mathbf{c}) - \sum_{\mathbf{c}} E[Y(G_{d|\mathbf{c}}(0), G_{m|\mathbf{c}}(0))|R = r, c]P(c), \text{ and} \\
\zeta(0) &\equiv \sum_{\mathbf{c}} E[Y(G_{d|\mathbf{c}}(0), G_{m|\mathbf{c}}(0))|R = r, c]P(c) - \sum_{\mathbf{c}} E[Y|R = 0, \mathbf{c}]P(\mathbf{c}),
\end{aligned}
\tag{1}
$$

where $r \in \{1, 2, 3\}$ and $\mathbf{d} \in \mathcal{D}, \mathbf{m} \in \mathcal{M}$, and $\mathbf{c} \in \mathcal{C}$. The disparity reduction $\delta(r)$ is the degree to which cardiovascular health for a comparison group ($R = r$) would change if the distributions of perceived discrimination and education were the same as those of White men, that is, how much the disparity would be reduced by the hypothetical intervention of equalizing distributions of perceived discrimination and education between the two groups. The disparity remaining is $\zeta(0)$, which is the degree to which the disparity for the comparison group ($R = r$) would remain if the distributions of the mediators were the same as those of White men. By combining the disparity reduction and disparity remaining, we can obtain the initial disparity as $\tau(r, 0) = \delta(r) + \zeta(0)$.

---

[1] The SUTVA assumes 1) that an individual does not affect the outcome of another individual and 2) that there is no variation in the treatment.

## 3.2. Identification Assumptions and Results

In this section, we present the identification assumptions and results for disparity reduction ($\delta(r)$) and disparity remaining ($\zeta(0)$). Three assumptions (A1-A3) that permit the identification are as follows.

**A1. Conditional ignorability**: $Y(d, m) \perp \{D, M\}|R = r, \mathbf{X} = \mathbf{x}, \mathbf{C} = \mathbf{c}$ for all $r \in \{0, 1, 2, 3\}$, $d \in \mathcal{D}, m \in \mathcal{M}, \mathbf{x} \in \mathcal{X}$, and $\mathbf{c} \in \mathcal{C}$.

**A2. Positivity and overlap**: $0 < P(R = r|\mathbf{c}) < 1$ and

$0 < P(D = d, M = m|R = r, \mathbf{X} = \mathbf{x}, \mathbf{C} = \mathbf{c})$ for all $r \in \{0, 1, 2, 3\}$, $\mathbf{x} \in \mathcal{X}$, $d \in \mathcal{D}, m \in \mathcal{M}$, and $\mathbf{c} \in \mathcal{C}$.

**A3. Consistency**: if $D_i = d$ and $M_i = m$ then $Y_i = Y(d, m)$ for all $d \in \mathcal{D}$, and $m \in \mathcal{M}$.

Conditional ignorability (A1) states that no unmeasured confounding exists between the outcome and the mediators jointly, given the intersectional group, intermediate confounders, and baseline covariates. Unlike sequential ignorability in the causal mediation literature (e.g., Imai et al., 2010; Pearl, 2001) based on natural indirect effects, this assumption is conditioned on post-exposure mediator and outcome confounding, which is $X_2$. This is an important advantage of using interventional effects instead of natural effects since it may not be plausible to assume the absence of post-exposure confounding in many settings. However, assumption A1 is still strong and will be violated in the presence of unmeasured mediator-outcome confounding. It is therefore essential to conduct sensitivity analysis that evaluates the robustness of the findings to potential violations of this assumption. Our second assumption (A2) implies that 1) the conditional probability of intersectional status as well as the conditional probability of mediators are positive (positivity), and 2) there is sufficient overlap in covariates across different race and gender combinations (region of common support). Our third assumption (A3) is that the observed outcome under a particular exposure value is the same as the outcome after intervening to set the exposure to that value (consistency).

Under assumptions A1-A3, the disparity reduction $(\delta(r))$ and remaining $(\zeta(0))$ are non-parametrically identified as

$$
\begin{aligned}
\delta(r) =& E[\frac{P(R=r)}{P(R=r|\mathbf{c})}y|R=r] - E[\frac{P(R=0)}{P(R=0|\mathbf{c})}\sum_{\mathbf{x}}E[Y|r,\mathbf{x},D,M,\mathbf{c}]P(\mathbf{x}|r,\mathbf{c})|R=0], \text{ and} \\
\zeta(0) =& E[\frac{P(R=0)}{P(R=0|\mathbf{c})}\sum_{\mathbf{x}}E[Y|r,\mathbf{x},D,M,\mathbf{c}]P(\mathbf{x}|r,\mathbf{c})|R=0] - E[\frac{P(R=0)}{P(R=0|\mathbf{c})}y|R=0],
\end{aligned}
\tag{2}
$$

where $r \in \{1,2,3\}$, $y \in \mathcal{Y}$, $\mathbf{x} \in \mathcal{X}$ and $\mathbf{c} \in \mathcal{C}$. A proof is given in appendix A.

## 4. Estimation

In this section, we propose an estimator that can effectively address multiple mediators and differential effects using imputation and inverse-probability scores. The estimator is built on the approach developed by VanderWeele and Vansteelandt (2014) for natural direct and indirect effects.

To calculate the initial disparity, we use regressions to model

$$
\hat{W}_r(\mathbf{C}_i) \equiv \frac{\hat{P}(R_i=r)}{\hat{P}(R_i=r|\mathbf{C}_i)},
\tag{3}
$$

where $r \in \{0,1,2,3\}$ and $c \in \mathcal{C}$. The probability of $R_i = r$ given covariates can be obtained by fitting a probit or logistic regression model. For instance, by fitting a multinomial logistic regression, $\hat{P}(R_i=1|\mathbf{C}_i) = \frac{exp(\hat{\lambda}_1\mathbf{C}_i)}{1+\sum_{r=1}^{3}exp(\hat{\lambda}_r\mathbf{C}_i)}$, where $\hat{\lambda}_r$s represent logit coefficients for $R = r$. A functional form fitted for the intersectional group given covariates should be correctly specified for a valid result. Using equation (3), the initial disparity between a comparison group $(R = r)$ and the reference group $(R = 0)$ can be estimated as

$$
\hat{\tau}(r,0) = \frac{1}{N_r}\sum_{i\in\Pi_r}\hat{W}_r(\mathbf{C}_i)Y_i - \frac{1}{N_0}\sum_{i\in\Pi_0}\hat{W}_0(\mathbf{C}_i)Y_i,
\tag{4}
$$

where $\Pi_r$ is the subjects (of size $N_r$) in intersectional group $R = r$ for $r \in \{0,1,2,3\}$.

To estimate disparity reduction/remaining, we follow the next three steps. First, we fit a model for intermediate confounders, regressing each confounder on the intersectional group and baseline covariates to the whole sample as $\psi_r(\mathbf{c}) \equiv P(\mathbf{X}_i|R_i=r,\mathbf{C}_i=\mathbf{c})$. Based on this fitted model, we compute a predicted confounder value for individual $i$ (denoted as $\tilde{\mathbf{x}}_i$) among the reference group after forcing $R$ to be equal to $r$.

Second, we fit an outcome model, regressing the outcome on the intersectional group, intermediate confounders, mediators, and baseline covariates to the whole sample as $\mu_{r\mathbf{x}dm}(\mathbf{c}) \equiv E[Y_i | R_i = r, \mathbf{X}_i = \mathbf{x}, D_i = d, M_i = m, \mathbf{C}_i = \mathbf{c}]$. Based on this fitted model, we compute a predicted outcome value for individual $i$ among the reference group after forcing $R$ to be equal to $r$ and imputing $\tilde{\mathbf{x}}_i$ as $\hat{\mu}_{r\tilde{\mathbf{x}}_i D_i, M_i}(\mathbf{C}_i)$. This computes a predicted estimate of the outcome for individual $i$ of the reference group, if the individual was in a comparison group but using the individual's own values of the mediators and baseline covariates.

Finally, $\hat{\mu}_{r\tilde{\mathbf{x}}_i D_i M_i}(\mathbf{C}_i)$ will be averaged over $i$ among the reference group given weight $\hat{W}_0(\mathbf{C}_i)$. Formally, the disparity reduction and disparity remaining are estimated, respectively, as

$$
\begin{aligned}
\hat{\delta}(r) &= \frac{1}{N_r} \sum_{i \in \Pi_r} \hat{W}_r(\mathbf{C}_i) Y_i - \frac{1}{N_0} \sum_{i \in \Pi_0} \hat{W}_0(\mathbf{C}_i) \hat{\mu}_{r\tilde{\mathbf{x}}_i D_i M_i}(\mathbf{C}_i), \quad \text{and} \\
\hat{\zeta}(0) &= \frac{1}{N_0} \sum_{i \in \Pi_0} \hat{W}_0(\mathbf{C}_i) \hat{\mu}_{r\tilde{\mathbf{x}}_i D_i M_i}(\mathbf{C}_i) - \frac{1}{N_0} \sum_{i \in \Pi_0} \hat{W}_0(\mathbf{C}_i) Y_i,
\end{aligned}
\tag{5}
$$

where $\Pi_r$ is the subjects (of size $N_r$) in intersectional group $R = r$ for $r \in \{0, 1, 2, 3\}$. The estimation requires several steps: calculating weights, calculating predicted values of $Y$ with incorporating a predicted value of $\mathbf{X}$, and calculating the weighted average. Therefore, we used nonparametric bootstrapping in order to obtain correct standard errors.

When fitting the outcome model, differential effects are assumed regarding perceived discrimination ($D$) across different intersectional groups, which is consistent with Bauer and Scheim (2019). Differential effects regarding education ($M$) across different intersectional groups can be easily specified, but we did not include it in our model because the interaction effect was not significant. For a valid result, the outcome and confounder models should be correctly specified. The estimate will be biased if differential effects are present but omitted from the outcome and confounder models.

As discussed above, this estimator is particularly useful when multiple mediators are considered because modeling mediators ($D$ and $M$) is not necessary. Specifying a correct functional form for multiple mediators can be challenging as the number of mediators increases. While this estimator is advantageous when more mediators exist than confounders, the advantage is lost when more confounders exist than mediators.

### An Aside: Regression Estimator

We review a regression estimator used by Jackson and VanderWeele (2018) to estimate the effects of interest (i.e., $\delta(r)$ and $\zeta(0)$) and to discuss strengths and weaknesses of the regression estimator. To begin, we assume the simplest model that does not assume any differential effects.

First, the initial disparity can be estimated by fitting the following regression:

$$Y = \phi_0 + \sum_{r=1}^{3} \phi_r I(R = r) + \phi_c \mathbf{C} + e_1, \tag{6}$$

where $I(R = r)$ is a dummy variable indicating $R = r$ for $r \in \{1, 2, 3\}$, and $e_1$ follows a normal distribution. Here, $\hat{\phi}_r$ would be the estimated observed disparity after conditioning on covariates.

Second, the disparity reduction and disparity remaining are estimated by fitting the following regressions:

$$\begin{aligned} Y =& \gamma_0 + \sum_{r=1}^{3} \gamma_r I(R = r) + \gamma_{\mathbf{x}} \mathbf{X} + \gamma_c \mathbf{C} + e_2, \text{ and} \\ Y =& \alpha_0 + \sum_{r=1}^{3} \alpha_r I(R = r) + \alpha_{\mathbf{x}} \mathbf{X} + \alpha_d D + \alpha_m M + \alpha_c \mathbf{C} + e_3, \end{aligned} \tag{7}$$

where $e_2$ and $e_3$ follow standard normal distributions. The term $\hat{\alpha}_r$ is the disparity remaining estimate after intervening on perceived discrimination and education ($D$ and $M$) within the same level of intermediate confounders ($\mathbf{X} = \mathbf{x}$). This estimand is not desirable because the disparity is estimated in the focus group in which the level of intermediate confounders (childhood SES and abuse) is the same, for instance, the group that had no exposure to child abuse. Also, this estimand prevents us from estimating a part of the disparity remaining, which is the path mediated via intermediate confounders (i.e., $R \to \mathbf{X} \to Y$) by conditioning on $\mathbf{X} = \mathbf{x}$. Therefore, to estimate the disparity remaining defined in equation (1), this path mediated via intermediate confounders $\mathbf{X}$ should be added, which is $\frac{\hat{\alpha}_{\mathbf{x}}}{\hat{\gamma}_{\mathbf{x}}} \cdot (\hat{\phi}_r - \hat{\gamma}_r)$. This is the disparity reduction when equalizing $\mathbf{X}$ alone but only scaled by the unmediated path between $\mathbf{X}$ and $Y$. Therefore, by combining these two effect estimates, $\zeta(0)$ is estimated as $\hat{\alpha}_r + \frac{\hat{\alpha}_{\mathbf{x}}}{\hat{\gamma}_{\mathbf{x}}} \cdot (\hat{\phi}_r - \hat{\gamma}_r)$.

The term $\hat{\gamma}_r - \hat{\alpha}_r$ is the disparity reduction estimate after intervening on perceived discrimination and education ($D$ and $M$) within the same level of $\mathbf{X} = \mathbf{x}$. However, the purpose is not to obtain disparity reduction within the same level of $\mathbf{X}$ and, thus, $\hat{\gamma}_r - \hat{\alpha}_r$ is insufficient to capture the disparity reduction. To obtain the disparity reduction defined in equation (1), we have to add $(1 - \frac{\hat{\alpha}_{\mathbf{x}}}{\hat{\gamma}_{\mathbf{x}}}) \cdot (\hat{\phi}_r - \hat{\gamma}_r)$, which is the disparity reduction when equalizing $\mathbf{X}$ alone but only scaled by the mediated path between $\mathbf{X}$ and cardiovascular health. By combining these two effect estimates, the disparity reduced is estimated as $\hat{\delta}(r) = \hat{\gamma}_r - \hat{\alpha}_r + (1 - \frac{\hat{\alpha}_{\mathbf{x}}}{\hat{\gamma}_{\mathbf{x}}}) \cdot (\hat{\phi}_r - \hat{\gamma}_r)$. For proofs, refer to Jackson and VanderWeele (2018).

The regression estimator yields in general efficient estimates, and the estimation is straightforward if the estimation is based on the subpopulation of $\mathbf{X} = \mathbf{x}$ (refer to the difference method in the structural equation model framework (Baron & Kenny, 1986; MacKinnon & Luecken, 2008). However, if the estimation is not based on the subpopulation, the regression estimator is no longer straightforward even after assuming the simplest model in which no differential effects exist. As models change, the disparity reduction and disparity remaining should be recalculated. This calls for an estimator that is suitable for a complex model with multiple mediators and possible differential effects such as the proposed estimator. Moreover, compared to the regression estimator, the proposed estimator can exploit the covariance balancing property of propensity scores (Imbens & Rubin, 2015), in which researchers can select a sample where the samples across intersectional groups are more balanced in terms of baseline covariates.

## 5. Sensitivity Analysis

Identification and estimation crucially rely on assumption A1, which is not empirically testable. This assumption requires no omitted confounding between the outcome and the two mediators jointly, given the variables measured before the mediators (i.e., the intersectional group, intermediate confounders, and baseline covariates). For this assumption to be met, even approximately, substantive knowledge is required about the confounding structure between the outcome and mediators. However, even with substantive knowledge, the assumption may be still violated because an important confounder identified based on substantive knowledge could be unmeasured/unmeasurable. To address possible violations of this assumption, we develop a sensitivity analysis that assesses the validity of an estimated disparity reduction (or remaining) to omitted mediator-outcome confounders. Built on VanderWeele (2010a), we present sensitivity analysis for multiple mediators based on regression coefficients.

We begin by calculating bias when unmeasured confounder $U$ exists that confounds the relationship between the mediators and the outcome. To do this, the following assumptions are required: 1) unmeasured confounder $U$ is independent with intersectional group $R$ given baseline covariates, and 2) unmeasured confounder $U$ is independent with intermediate confounders $\mathbf{X}$ given baseline covariates. These assumptions are met in Figure 2 but would be violated if $U$ is correlated with $R$ or $\mathbf{X}$ given baseline covariates. These assumptions are restrictive but are reasonably met in some settings, for example, if the unmeasured confounder is an unknown genetic factor that does not affect childhood experiences but affects later education and cardiovascular health.

Suppose that the disparity reduction for a comparison group ($R = r$) is estimated given measured variables as, $\delta(r) = \sum_{\mathbf{c}} E[Y|R = r, \mathbf{c}]P(\mathbf{c}) - \sum_{\mathbf{x},d,m,\mathbf{c}} E[Y|R = r, \mathbf{x}, d, m, \mathbf{c}]P(\mathbf{X} = \mathbf{x}|R = r, \mathbf{c})P(D = d, M = m|R = 0, \mathbf{c})P(\mathbf{c})$, using the identification result shown in appendix A (see equation (12)). If unmeasured confounder $U$ exists, this expression will lead to a biased estimate. We define this bias as the difference between the

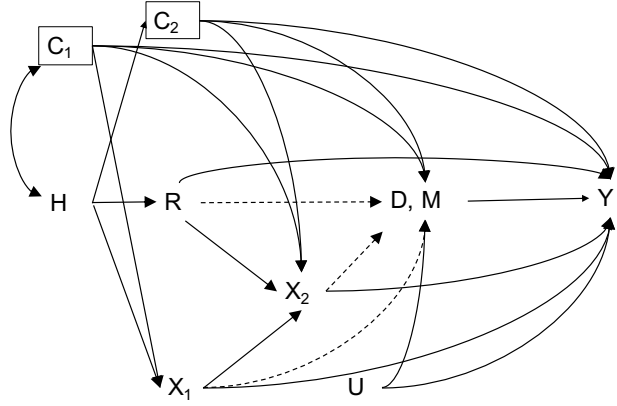

*Figure 2*. When unmeasured confounder $U$ between the mediators and outcome exists

Note. 1) Diagram represents the relationship between race and gender intersectional status $R$, cardiovascular health $Y$, discrimination $D$, and education $M$, as well as history $H$, age $C_1$, genetic vulnerability $C_2$, childhood SES $X_1$, and child abuse $X_2$.
2) Solid lines represent relationships that are preserved and dashed lines represent relationships that are removed by intervening on $D$ and $M$.
3) Placing a box around the conditioning variables implies that a disparity is considered within levels of these variables.

expected value of the estimate and the true effect. Formally, the bias for disparity reduction for $R = r$ $(bias(\delta(r)))$ is defined as

$$\sum_{\mathbf{c}} E[Y|R=r,\mathbf{c}]P(\mathbf{c}) - \sum_{\mathbf{x},d,m,\mathbf{c}} E[Y|R=r,\mathbf{x},d,m,\mathbf{c}]P(\mathbf{x}|R=r,\mathbf{c})P(d,m|R=0,\mathbf{c})P(\mathbf{c})$$

$$-\sum_{\mathbf{c}} E[Y|R=r,\mathbf{c}]P(\mathbf{c}) + \sum_{\mathbf{x},d,m,\mathbf{c},u} E[Y|R=r,\mathbf{x},d,m,\mathbf{c},u]P(\mathbf{x}|R=r,\mathbf{c})P(d,m|R=0,\mathbf{c},u)P(u,\mathbf{c}).$$

Given this definition, general bias formulas for disparity reduction and remaining can be calculated (refer to appendix B); yet, we only present the simplified bias formulas that are more practical. The general bias formulas are simplified under the following assumptions: 1) unmeasured confounder $U$ is binary, 2) the effect of unmeasured confounder $U$ on the outcome is constant across the strata of the intersectional group $(R)$, intermediate confounders $(\mathbf{X})$, mediators $(D, M)$, and baseline covariates $(\mathbf{C})$, and 3) the effect of the intersectional group on the prevalence of unmeasured confounder $U$ is constant across the strata of mediators $(D, M)$ and baseline covariates $(\mathbf{C})$. These simplifying assumptions may be violated in some applied settings and, thus we will discuss a way to relax these assumptions in section 5.2.

## 5.1. Sensitivity Analysis Based on the Regression Coefficients

One way to assess the sensitivity of an estimated disparity reduction (or remaining) is to use the following sensitivity parameters: 1) the difference in expected outcome $Y$ comparing $U = 1$ and $U = 0$ is $\gamma$, across strata of $R, \mathbf{X}, D, M$, and $\mathbf{C}$, and 2) the prevalence of unmeasured confounder $U$ comparing intersectional statuses $R = r$ and $R = 0$ is $\beta_r$ across strata of $D, M$, and $\mathbf{C}$. We provide the simplified bias formulas given these two sensitivity parameters as

$$
\begin{aligned}
bias(\delta(r)) &= -\gamma\beta_r, \text{ and} \\
bias(\zeta(0)) &= \gamma\beta_r,
\end{aligned}
\tag{8}
$$

where $r \in \{1, 2, 3\}$. A proof is given in appendix C. From the bias formulas, we note that the bias for disparity reduction and disparity remaining is the same except that the signs are the opposite. Unmeasured confounder $U$ is assumed to be independent with the race-gender intersectional status given baseline covariates, and thus the bias for the initial disparity due to the unmeasured confounder is zero.

While the interpretation of $\gamma$ is straightforward, the interpretation of $\beta_r$ may not be straightforward. The sensitivity parameter ($\beta_r$) is conditioned on the mediators, which are descendants (variables that are affected by) of unmeasured confounder $U$. The prevalence difference in unmeasured confounder $U$ between the marginalized and non-marginalized groups may be zero but, the difference could be negative or positive after conditioning on mediators. For instance, suppose that the prevalence of a genetic factor ($U$) between Black women and White men is equal; suppose further that discrimination ($D$) occurs if either a person is Black women or has the genetic factor. If a White man is discriminated against, then we know that he has the genetic factor. Thus, by conditioning on discrimination, the genetic factor becomes more prevalent for White men than Black women (See, e.g.,

VanderWeele, 2010b; VanderWeele & Robins, 2007, for more explanation). While the interpretation of $\beta_r$ is challenging, existing covariates could give a range of plausible values of this sensitivity parameter.

Note that $\beta_r$ does not condition on intermediate confounders $\mathbf{X}$, which fixes the error in Park and Esterling (2020). The bias includes the path of $R \to (\mathbf{X}) \to D, M \to U \to Y$ where the parentheses imply that the path goes both through or not through $\mathbf{X}$.

## 5.2. When a Simplifying Assumption Is Violated

When one of the simplifying assumptions is violated, one can use the original bias formula shown in appendix B (See, equation (14)), which may not be very practical since the number of sensitivity parameters becomes unwieldy. Therefore, we recommend modifying the original bias formula depending on the particular violation(s). Suppose that the second simplifying assumption is violated such that the effect of the unmeasured confounder on the outcome depends on the value of mediators as $\gamma^{\mathbf{dm}}$; and that the third assumption is met that the prevalence difference in the unmeasured confounder is constant across the level of mediators and covariates as $\beta_r$. Then the following modified bias formula can be used as an alternative.

$$bias(\delta(r)) = \beta_r \sum_{d,m,c} \gamma^{dm} P(d,m|R=0,\mathbf{c})P(\mathbf{c}), \tag{9}$$

where $d \in \mathcal{D}, m \in \mathcal{M}$, and $\mathbf{c} \in \mathcal{C}$. The conditional probability of mediators (i.e., $P(d,m|R=0,\mathbf{c})$) can be drawn from the data. Then the overall value of $\gamma$ will be obtained by summing over values of $\gamma^{dm}$ weighted by this conditional probability of mediators.

As another example, suppose that the second simplifying assumption is met while the third assumption is violated such that the prevalence difference in the unmeasured confounder depends on the value of baseline covariates as $\beta_r^{\mathbf{c}}$. Then the following modified bias formula can be used as

$$bias(\delta(r)) = \gamma \sum_c \beta_r^{\mathbf{c}} P(\mathbf{c}), \tag{10}$$

where $\mathbf{c} \in \mathcal{C}$. Again, the overall value of $\beta_r$ will be obtained by summing over values of $\beta_r^{\mathbf{c}}$ weighted by the distribution of covariates (i.e., $P(\mathbf{c})$).

## 6. Application to the MIDUS Data

### 6.1. Data and Measures

Extending analyses by Lee, Park, and Boylan (2020), we extracted baseline and outcome data from MIDUS and the MIDUS Refresher. We limited the sample to those respondents (n=1978) who participated in MIDUS wave 2 or MIDUS Refresher biological data collection and identified themselves either as non-Hispanic White or non-Hispanic Black. Racial and gender statuses were created using the nexus of self-identified race/ethnicity and gender. Cardiovascular health was assessed in accordance with the AHA's criteria (Lloyd-Jones et al., 2010). A composite was created to reflect the criteria for ideal, intermediate, or poor cardiovascular health, respectively, on each of seven metrics: smoking, BMI, physical activity, diet, total cholesterol, blood pressure, and fasting glucose.

We have considered two life-course mediators (perceived discrimination and education) that explain cardiovascular disparities across intersectional groups. As for perceived discrimination, respondents were asked to report the number of times in their life they faced "discrimination" in 11 questions. Each item was recoded 1 if respondents reported 1 or more times, otherwise 0. An inventory of lifetime discrimination was constructed by summing the items with possible scores ranging from 0 to 11 (Williams et al. 1997). Education is a variable that indicates the highest level of degree completed, which ranges from 1 = no school/some grade school to 12 = PhD, MD, or other professional degree.

Intermediate confounders include childhood SES and abuse. Childhood SES is an index measure including parental education, poverty, financial status, and employment status of parent(s). Abuse is an index, drawn from items on the Childhood Trauma Questionnaire (Bernstein & Fink, 1998), measuring experiences of emotional, physical, or sexual abuse, with possible responses to each item ranging from 1 (never true) to 5 (very often true).

For baseline covariates, we included age and parental history of cardiovascular and metabolic illness (heart problems, stroke, and diabetes), which may reflect genetic susceptibility and shared lifestyle/environments associated with reduced respondent's cardiovascular health score.

Consistent with Lee et al. (2020), while Black women in our study have the lowest (i.e., unhealthiest) cardiovascular health scores (6.99), White women have the highest cardiovascular health scores (8.73). White men have higher scores than Black men (7.95 vs. 7.29), and there is no significant gender difference among Blacks. Regardless of gender, Black adults show greater levels of standardized perceived discrimination scores (Black men=0.81, Black women=0.69) than Whites (White men= -0.28, White women= -0.13). White men have the highest levels of standardized education scores (0.37), followed by White women (0.27), Black women (-0.31) and Black men (-0.40), although there is no significant gender difference among Blacks.

## 6.2. Analysis under Conditional Ignorability

We compared the results from both proposed and regression-based methods. For the proposed method, we used the estimator shown in equation (5) and calculated the estimates with and without considering differential effects across intersectional groups. For the regression-based method, we used the regression estimator suggested by Jackson and VanderWeele (2018) and calculated the estimates without considering differential effects. It is possible to consider differential effects with the regression estimator, but this requires additional calculations.

The results of all comparison groups (Black men, White women, and Black women)

are available, but for simplicity, we only present the disparity reduction and disparity remaining for Black women when compared to White men after intervening on education and perceived discrimination simultaneously.

Table 1

*Estimates of the disparity reduction and disparity remaining for Black women vs. White men*

| Estimator | Proposed Estimator | Regression Estimator |
|---|---|---|
| Observed disparity ($\tau(3,0)$) | -0.987 | -0.927 |
| (95% CI) | (-1.281, -0.684) | (-1.240, -0.617) |
| Without Differential Effects | | |
| Disparity remaining ($\zeta(0)$) | -0.401 | -0.396 |
| (95% CI) | (-0.693, -0.065) | (-0.704 -0.072) |
| Disparity reduction ($\delta(3)$) | -0.586 | -0.531 |
| (95% CI) | (-0.776, -0.392) | (-0.688 -0.386) |
| % reduction | 59.4% | 57.3% |
| With Differential Effects | | |
| Disparity remaining ($\zeta(0)$) | -0.524 | |
| (95% CI) | ( -0.883, -0.193) | |
| Disparity reduction ($\delta(3)$) | -0.464 | |
| (95% CI) | (-0.701, -0.227) | |
| % reduction | 47.0% | |

*Note.* CI = confidence interval.

Table 1 shows initial disparities ($\tau(3,0)$), disparity remaining ($\zeta(0)$), and disparity reduction ($\delta(3)$) in terms of cardiovascular health. The results from the proposed estimator show that, compared to White men, the initial disparity for Black women is -0.987. Similarly, the results from regression-based methods show that, compared to White men, the initial disparity for Black women is -0.927. The difference is probably due to the following factors: 1) the proposed estimator requires the covariate balancing property of propensity scores, and 2) the proposed estimator standardizes covariates across intersectional groups rather than conditioning on them. Both estimates are negative, with a confidence interval bounded away from zero, which means that cardiovascular health for Black women is significantly lower than White men among those who have the same distributions of age and genetic vulnerability. While this result is interesting, sociologists have a further interest in knowing which interventions are necessary to reduce this disparity.

The second two rows show disparity remaining and reduction without specifying differential effects for discrimination. This implies that the effect of discrimination on cardiovascular health is assumed to be constant across intersectional groups. After equalizing the distributions of both education and perceived discrimination across groups, the results from the proposed method show that the initial disparity would be reduced by 59.4% for Black women compared to White men. The regression-based method results show comparable levels of disparity reduction for Black women (57.3%) compared to White men.

The third two rows show disparity remaining and reduction when considering differential effects for discrimination. As we consider whether the effect of perceived discrimination on cardiovascular health varies by intersectional group, the results from the proposed method show that the initial disparity would be reduced by 47.0% for Black women. Note that the percentage of reduction when considering differential effects decreased by 12.4%. That's because the effect of discrimination on cardiovascular health is larger for White men than Black women, which is consistent with prior work that shows a salient inverse association between interpersonal discrimination and cardiovascular health scores, particularly for White adults (Bey, Jesdale, Forrester, Person, & Kiefe, 2019). This finding highlights the importance of addressing differential effects. Overall, the results suggest that cardiovascular health disparities for Black women will be significantly reduced if interventions could decrease the level of discrimination while simultaneously increasing the level of education to that of White men with the same age and genetic vulnerability distributions.

### 6.3. Sensitivity Analysis

In this section, we apply the sensitivity analyses that we derived from the bias formula results of Section 5.1. The sensitivity of the results depends on the following two parameters: 1) the effect of an unmeasured confounder on the cardiovascular health given the intersectional status, intermediate confounders, mediators, and baseline covariates ($\gamma$) and 2) the prevalence difference in an unmeasured confounder comparing Black women and White men given mediators and baseline covariates ($\beta_r$). Under the simplifying assumptions described above, the bias for disparity reduction is $-\gamma\beta_r$, and the bias for disparity remaining is $\gamma\beta_r$.

We use existing covariates as reference values to reasonably speculate a plausible range of each parameter's magnitude. The parameter $\gamma$ ranges from -0.584 (age) to -0.098 (parental history of stroke). The parameter $\beta_r$ ranges from -0.157 (parental history of heart diseases) to 0.204 (parental history of diabetes). While these reference values lend context to the parameters, it is essential to note that an unmeasured confounder could exist with a larger association with the outcome and the exposure than these values. For this illustration, let us assume unmeasured confounders act toward reducing the absolute value of disparity reduction/remaining.

Figure 3A indicates that the disparity reduction is robust to violations of the no omitted mediator-outcome confounding assumption. If we have an unmeasured confounder as strong as age ($\gamma \approx 0.584$), the estimate for disparity reduction will become zero with the value of $\beta_r = 0.795$ (because $-0.464 - (-0.584 \times 0.795) \approx 0$). With the same strength of an unmeasured confounder, the 95% confidence interval will cover zero with the value of $\beta_r = 0.389$ (because $-0.227 - (-0.584 \times 0.389) \approx 0$). In other words, the prevalence difference in an unmeasured confounder comparing Black women and White men should be at least 38.9% after conditioning on mediators for the disparity reduction to change from significance to non-significance at the 95% confidence level. This amount of confounding is greater than the range of reference values for $\beta_r$ drawn from existing covariates.

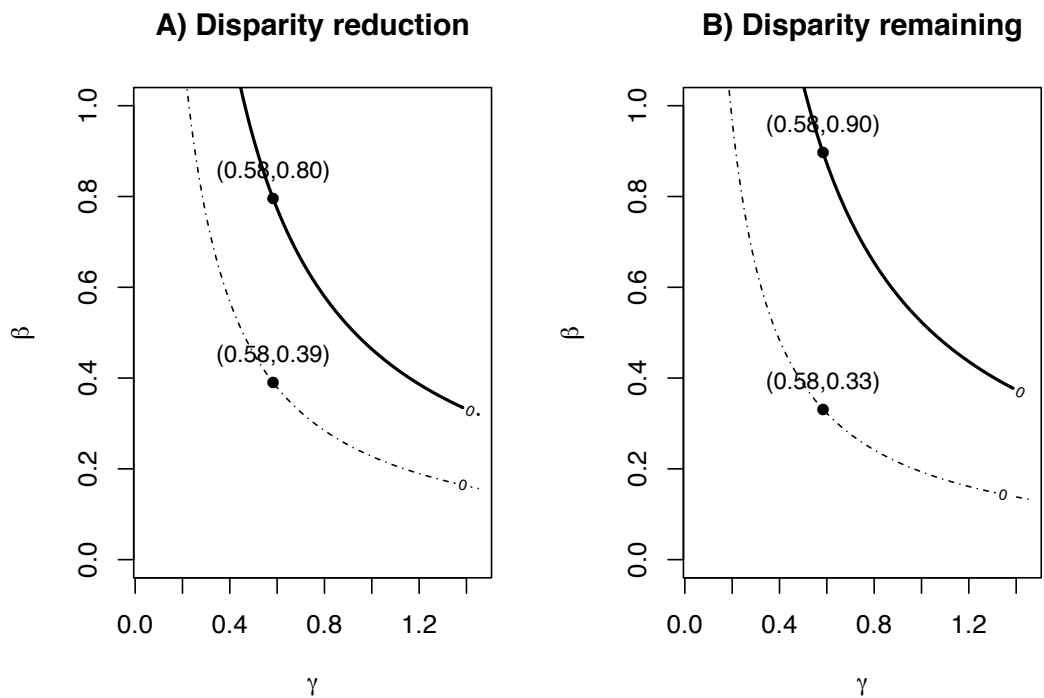

*Figure 3.* Sensitivity of the estimates based on regression coefficients

*Note. 1) Bold lines represent the points at which the estimates become zero. 2) Dashed lines represent the points at which the 95% confidence intervals cover zero. 3) The estimates' standard errors are assumed to be invariant with a varying amount of confounding.*

The same applies to the disparity remaining estimate. Figure 3B indicates that the disparity remaining is robust to violations of the no omitted mediator-outcome confounding assumption. The estimate for disparity remaining will become zero with the value of $\beta_r = 0.897$ (because $-0.524 - (-0.584 \times 0.897) \approx 0$); the 95% confidence interval will cover zero with the value of $\beta_r = 0.330$ (because $-0.193 - (-0.584 \times 0.330) \approx 0$); given that we have an unmeasured confounder as strong as age. This confounding amount is still greater than the reference values drawn from existing covariates. This result implies that the significance of disparity remaining will not change when there are unmeasured confounders with a similar range of sensitivity parameter values as the existing covariates.

## 7. Discussion

In this paper, we study identification, estimation, and sensitivity analysis for the disparity reduction and disparity remaining between intersectional groups. Our paper contributes to the causal decomposition analysis literature in several ways. First, we developed a nonparametric estimator, which effectively accommodates a complex model with multiple mediators and differential effects. Second, we developed a sensitivity analysis based on the regression coefficients. The proposed sensitivity analysis technique can help researchers assess their findings' robustness to possible violations of the no omitted mediator-outcome confounding assumption even when the model underlying their study is complex.

The recent work by Jackson (2020) proposed alternative weighting methods (i.e., the ratio of mediator probability and inverse odds ratio weightings) in disparity research that can be used when a single mediator is considered. These estimators are also flexible to accommodate differential effects and can be modified to accommodate multiple mediators. They require a functional form to be correctly specified for every intervening mediator and intersectional group. In contrast, the proposed estimator requires a functional form to be correctly specified for the outcome, intermediate confounders, and intersectional group. Researchers can choose between these estimators depending on the context of their studies; yet, the proposed estimator is less demanding in terms of modeling perspectives as the number of mediators increases.

While the primary purpose of this paper is methodological, we note issues related to perceived discrimination. Our measure of discrimination mainly captures individuals' awareness of or willingness to report discrimination. It is possible that experiencing discrimination might have different meanings or that different reporting thresholds exist across intersectional groups. For example, White men might view discrimination as a loss of white supremacy or privilege, while Black women might encounter multiple forms of discrimination through racial or male supremacy. If a differential construct is used as a mediator, it is difficult to interpret the disparity reduction/remaining in a meaningful way since it is not clear how to equalize, even hypothetically, the distribution of perceived discrimination across groups. Jackson and VanderWeele (2019) discussed this problem of differential constructs and offered several solutions.

There are limitations and possible future directions for the proposed methods. First, the proposed estimator is developed for interventions that depend on baseline covariates only. For example, we intervene so that the distributions of education and discrimination were the same between groups who have the same age and genetic vulnerability. However, suppose investigators are interested in other types of interventions, such as equalizing the mediators' distributions between groups who have the same early life adversities (i.e., child abuse or child SES) in addition to the baseline covariates. In that case, a different estimator should be considered. For more explanation, refer to Jackson (2020). Second, it is important to note that the proposed sensitivity analysis is restricted to certain settings due to its assumptions. For example, unmeasured confounders would likely be correlated with the intersectional group and intermediate confounders in many situations. Therefore, an important area for future study is to develop a general sensitivity analysis that does not require these assumptions. Third, another critical area for future study is to address possible measurement errors in confounders and mediators. Given that measurement error for confounders and mediators is common, it would be necessary to develop sensitivity analysis to check the robustness of the results to possible measurement error in the context of disparity research.

## Appendix A: Identification of $\delta(r)$ and $\zeta(0)$

As defined in equation (1), $\delta(r) = \sum_{\mathbf{c}} E[Y|r, \mathbf{c}]P(\mathbf{c}) - E[Y(G_{d|c}(0), G_{m|c}(0))|r]$, and $\zeta(0) = E[Y(G_{d|c}(0), G_{m|c}(0))|r] - \sum_{\mathbf{c}} E[Y|R = 0, \mathbf{c}]P(\mathbf{c})$. First, we can express $\sum_{c} E[Y|r, \mathbf{c}]P(\mathbf{c})$ using weight as

$$
\begin{aligned}
\sum_{c} E[Y|r, \mathbf{c}]P(\mathbf{c}) &= \sum_{\mathbf{c}, y} y P(y|r, \mathbf{c})P(\mathbf{c}) \\
&= \sum_{\mathbf{c}, y} \frac{P(r)}{P(r|\mathbf{c})} y P(y|r, \mathbf{c})P(r|\mathbf{c})P(\mathbf{c}) \frac{1}{P(r)} \\
&= \sum_{\mathbf{c}, y} \frac{P(r)}{P(r|\mathbf{c})} y P(y, \mathbf{c}|r) \\
&= E[\frac{P(r)}{P(r|\mathbf{c})} y | r].
\end{aligned}
\tag{11}
$$

The first equality is due to the law of iterated expectations. The third equality is due to Bayes theorem.

Under assumptions A1-A3, the counterfactual $E[Y(G_{d|\mathbf{c}}(0), G_{m|\mathbf{c}}(0))|R = r]$ can be identified as

$$
\begin{aligned}
&= \sum_{\mathbf{c}} E[Y(G_{d|\mathbf{c}}(0), G_{m|\mathbf{c}}(0))|r, \mathbf{c}]P(\mathbf{c}) \\
&= \sum_{d, m, \mathbf{c}} E[Y(d, m)|r, G_{d|\mathbf{c}}(0) = d, G_{m|\mathbf{c}}(0) = m, \mathbf{c}]P(G_{d|\mathbf{c}}(0) = d, G_{m|\mathbf{c}}(0) = m|r, \mathbf{c})P(\mathbf{c}) \\
&= \sum_{d, m, \mathbf{c}} E[Y(d, m)|r, \mathbf{c}]P(d, m|R = 0, \mathbf{c})P(\mathbf{c}) \\
&= \sum_{d, m, \mathbf{c}, \mathbf{x}} E[Y(d, m)|r, \mathbf{x}, \mathbf{c}]P(\mathbf{x}|r, \mathbf{c})P(d, m|R = 0, \mathbf{c})P(\mathbf{c}) \\
&= \sum_{d, m, \mathbf{c}, \mathbf{x}} E[Y(d, m)|r, \mathbf{x}, d, m, \mathbf{c}]P(\mathbf{x}|r, \mathbf{c})P(d, m|R = 0, \mathbf{c})P(\mathbf{c}) \\
&= \sum_{d, m, \mathbf{c}, \mathbf{x}} E[Y|r, \mathbf{x}, d, m, \mathbf{c}]P(\mathbf{x}|r, \mathbf{c})P(d, m|R = 0, \mathbf{c})P(\mathbf{c}).
\end{aligned}
\tag{12}
$$

The first, second, and fourth equalities are due to the law of iterated expectations. The third equality holds because $G_{d|c}(0) = d$, and $G_{m|c}(0) = m$ are random given $\mathbf{C} = \mathbf{c}$. The fifth equality holds due to assumption A1, $(Y(d, m) \perp \{D, M\}|r, \mathbf{X} = \mathbf{x}, \mathbf{C} = \mathbf{c})$. The last equality holds due to A3 (consistency). Then, the last expression of equation (12) equals

$$
\begin{aligned}
&= \sum_{\mathbf{c},\mathbf{x},D,M,R} I(R=0)E[Y|r,\mathbf{x},D,M,\mathbf{c}]P(\mathbf{x}|r,\mathbf{c})P(D,M|R,\mathbf{c})P(\mathbf{c}) \\
&= \sum_{\mathbf{c},\mathbf{x},D,M,R} \frac{I(R=0)}{P(R=0|\mathbf{c})}E[Y|r,\mathbf{x},D,M,\mathbf{c}]P(\mathbf{x}|r,\mathbf{c})P(D,M|R,\mathbf{c})P(R|\mathbf{c})P(\mathbf{c}) \\
&= \sum_{\mathbf{c},\mathbf{x},D,M,R} \frac{I(R=0)}{P(R=0|\mathbf{c})}E[Y|r,\mathbf{x},D,M,\mathbf{c}]P(\mathbf{x}|r,\mathbf{c})P(D,M,R,\mathbf{c}) \\
&= E[\frac{I(R=0)}{P(R=0|\mathbf{c})}\sum_{\mathbf{x}}E[Y|r,\mathbf{x},D,M,\mathbf{c}]P(\mathbf{x}|r,\mathbf{c})] \\
&= E[\frac{P(R=0)}{P(R=0|\mathbf{c})}\sum_{\mathbf{x}}E[Y|r,\mathbf{x},D,M,\mathbf{c}]P(\mathbf{x}|r,\mathbf{c})|R=0].
\end{aligned}
\tag{13}
$$

In the first equality, we use $D, M$, and $R$ to represent that these are random variables. The fourth equality holds because $\sum_b E[a|b]P(b) = E[E[a|b]]$. This completes the proof.

## Appendix B: General Bias Formulas of $\delta(r)$ and $\zeta(0)$

The bias for disparity reduction for $R = r$ (bias($\delta(r)$)) is defined as the difference between the expected estimate and the true value as

$$\sum_c E[Y|R = r, c]P(c) - \sum_{x,d,m,c} E[Y|R = r, x, d, m, c]P(x|R = r, c)P(d, m|R = 0, c)P(c) -$$

$$\sum_c E[Y|R = r, c]P(c) + \sum_{x,d,m,c,u} E[Y|R = r, x, d, m, c, u]P(x|R = r, c)P(d, m|R = 0, c, u)P(c, u)$$

$$= -\sum_{x,d,m,c} E[Y|R = r, x, d, m, c]P(u|r, x, d, m, c)P(x|R = r, c)P(d, m|R = 0, c)P(c)$$

$$+ \sum_{x,d,m,c,u} E[Y|R = r, x, d, m, c, u]P(x|R = r, c)P(d, m|R = 0, c, u)P(c, u)$$

$$= -\sum_{x,d,m,c,u} E[Y|R = r, x, d, m, c, u]P(u|R = r, x, d, m, c)P(x|R = r, c)P(d, m|R = 0, c)P(c)$$

$$+ \sum_{x,d,m,c,u} E[Y|R = r, x, d, m, c, u]P(x|R = r, c)\frac{P(u|R = 0, d, m, c)}{P(u|R = 0, c)}P(d, m|R = 0, c)P(u|c)P(c)$$

$$= -\sum_{x,d,m,c,u} E[Y|R = r, x, d, m, c, u]\{P(u|R = r, x, d, m, c) - P(u|R = 0, d, m, c)\}$$

$$\times P(d, m|R = 0, c)P(x|R = r, c)P(c)$$

$$= -\sum_{x,d,m,c,u} E[Y|R = r, x, D(0) = d, M(0) = m, c, u]\{P(u|R = r, x, D(0) = d, M(0) = m, c)$$

$$- P(u|R = 0, D(0) = d, M(0) = m, c)\}P(D(0) = d, M(0) = m|R = 0, c)P(x|R = r, c)P(c)$$

$$= -\sum_{x,d,m,c,u} E[Y|R = r, x, D(0) = d, M(0) = m, c, u]\{P(u|R = r, D(0) = d, M(0) = m, c)$$

$$- P(u|R = 0, D(0) = d, M(0) = m, c)\}P(D(0) = d, M(0) = m|R = 0, c)P(x|R = r, c)P(c)$$

$$= -\sum_{x,d,m,c,u} E[Y|R = r, x, d, m, c, u]\{P(u|R = r, d, m, c) - P(u|R = 0, d, m, c)\}$$

$$\times P(d, m|R = 0, c)P(x|R = r, c)P(c).$$

The first equality is due to the law of iterated expectations. The second equality is due to Bayes theorem. The third equality is because $U$ only confounds the relationship between the mediators and outcome (and thus, $U \perp R = r|c$). The fourth equality is due to A3 (consistency). The fifth equality holds due to $U \perp X|R = r, D(0) = d, M(0) = m, C = c$, which is met since $D(0)$ and $M(0)$ are only affected by $R = 0$ but not $X(r)$. The last equality is due to A3 (consistency).

Then, for reference value of $U = u'$, we have

$$= -\sum_{x,d,m,c,u} \{E[Y|R=r,x,d,m,c,u] - E[Y|R=r,x,d,m,c,u']\}\{P(u|R=r,d,m,c) - P(u|R=0,d,m,c)\}$$

$$\times P(d,m|R=0,c)P(x|R=r,c)P(c).$$

$$(14)$$

The equality is due to $\sum_u E[Y|R=r,x,d,m,c,u']P(u|R=r,d,m,c) = \sum_u E[Y|R=r,x,d,m,c,u']P(u|R=0,d,m,c)$. This is the general bias formula for $\delta(r)$. Since $U \perp R = r|C = c$, the bias for the observed disparity due to $U$ is zero. Therefore, bias($\zeta(0)$)= -bias($\delta(r)$). This completes the proof.

**Appendix C: Simple Bias Formulas Based on the Regression Coefficients**

Suppose that 1) $U$ is binary, 2) the difference in expected outcome $Y$ comparing U=1 and U=0 is constant across the strata of $R, X, D, M$, and $C$, and 3) the prevalence of unmeasured confounder $U$ comparing intersectional statuses $R = r$ and $R = 0$ is constant across the strata of $D, M$, and $C$. Then, equation (14) reduces to

$$
\begin{aligned}
= & -\sum_{x,d,m,c} \{E[Y|R = r, x, d, m, c, U = 1] - E[Y|R = r, x, d, m, c, U = 0]\} \\
& \times \{P(U = 1|R = r, d, m, c) - P(U = 1|R = 0, d, m, c)\}P(d, m|R = 0, c)P(x|R = r, c)P(c) \\
= & -\{E[Y|R = r, x, d, m, c, U = 1] - E[Y|R = r, x, d, m, c, U = 0]\} \\
& \times \{P(U = 1|R = r, d, m, c) - P(U = 1|R = 0, d, m, c)\}. \\
= & -\gamma\beta_r,
\end{aligned}
\tag{15}
$$

where $\gamma = E[Y|R = r, x, d, m, c, U = 1] - E[Y|R = r, x, d, m, c, U = 0]$ and $\beta_r = P(U = 1|R = r, d, m, c) - P(U = 1|R = 0, d, m, c)$. This completes the proof.

## Author Note

The MIDAS data used for the case study can be accessed via ICPSR (https://www.icpsr.umich.edu/web/ICPSR/series/203) or the MIDUS Colectica Portal (https://midus.colectica.org/Account/Login). The R code for implementing analytic approaches proposed in this article can be found in appendix D. In the R code, we used the "mmi" function from the R package *causal.decomp* (Kang & Park, 2021) to estimate initial disparity, disparity reduction, and disparity remaining.